\documentclass[aps,pra,reprint,groupedaddress]{revtex4-1}

\usepackage{amsmath}
\usepackage{amssymb}
\usepackage{graphicx}
\usepackage{verbatim}
\usepackage{lineno}

\pdfoutput=1

\newcommand{\eq}{\begin{equation}}
\newcommand{\eeq}{\end{equation}}
\newcommand{\ket}[1]{\left|#1\right\rangle}
\newcommand{\bra}[1]{\left\langle #1\right|}

\begin{document}

\title{Quantum Gates with Phase Stability over Space and Time}

\author{I. V. Inlek}
\email{inlek@umd.edu}
\author{G. Vittorini}
\author{D. Hucul}
\author{C. Crocker}
\author{C. Monroe}
\affiliation{Joint Quantum Institute, University of Maryland Department of Physics and National Institute of Standards and Technology, College Park, Maryland 20742}

\date{\today}

\begin{abstract}

The performance of a quantum information processor depends on the precise control of phases introduced into the system during quantum gate operations. As the number of operations increases with the complexity of a computation, the phases of gates at different locations and different times must be controlled, which can be challenging for optically-driven operations.  We circumvent this issue by demonstrating an entangling gate between two trapped atomic ions that is insensitive to the optical phases of the driving fields, while using a common master reference clock for all coherent qubit operations.  Such techniques may be crucial for scaling to large quantum information processors in many physical platforms.

\end{abstract}

\maketitle

\section{INTRODUCTION}
In a quantum information processor, the control and entanglement of quantum bits is usually accomplished with external electromagnetic fields, whose phase is directly imprinted on the qubits \cite{quantumcomputers}.  Generating large-scale entanglement for applications in quantum information science therefore relies upon the spatial and temporal coherence of phases throughout the system.  As the system grows in complexity to many qubits and quantum gate operations, likely requiring a modular architecture \cite{musiqc}, it will become crucial to control and coordinate the phases between modules and between qubits within a module.

In this paper, we demonstrate the absolute control of qubit phases in both space and time using a collection of trapped atomic ion qubits driven by optical fields.  We choose appropriate beam geometries that eliminate the dependence of qubit phases on absolute optical path lengths from the driving field, and we use a common high quality master oscillator as a reference for all operations.  These techniques are applicable to many other quantum computing platforms such as NV-centers in diamond \cite{nv_centers}, optical quantum dots \cite{qdots}, and optical lattices containing neutral atoms \cite{neutrals}.

\section{EXPERIMENTAL RESULTS}
\subsection{Single-qubit gates}
We consider qubit states with rf or microwave frequency splittings, as opposed to optical qubit splittings which require absolute optical phase stability \cite{optical_qubit}. We use qubits encoded in the hyperfine clock states of trapped ${}^{171}$Yb$^{+}$ atoms $\ket{F = 0, m_F = 0} \equiv \ket{0}$ and $\ket{F = 1, m_F = 0} \equiv \ket{1}$ of the ${}^2S_{1/2}$ manifold with a hyperfine splitting of $\omega_0/2\pi = \nu_0 = 12.64282$ GHz. Standard photon scattering methods are used for Doppler cooling, state initialization and detection \cite{ybqubit}.

\begin{figure}
\includegraphics[width = 3.3 in]{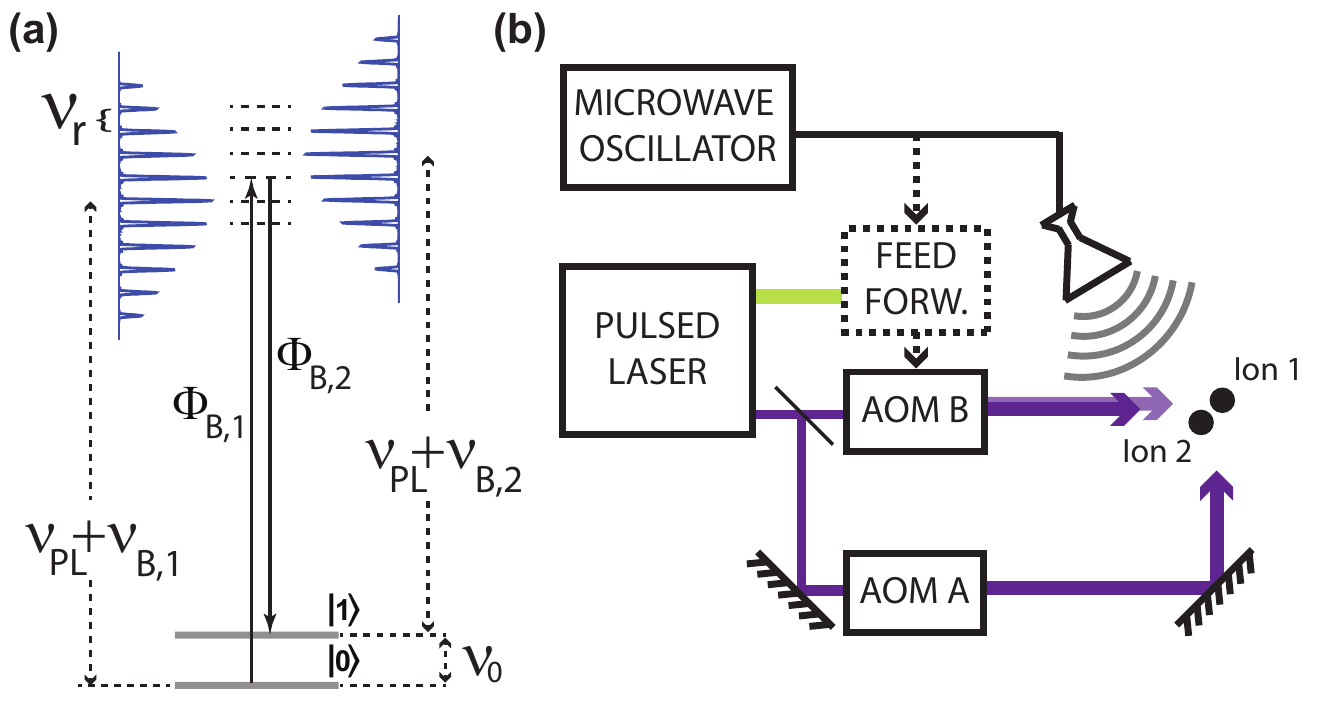}
\caption{(color online). (a) The qubit is driven from atomic levels $\ket{0}$ to $\ket{1}$ via two-photon stimulated Raman process by absorbing from the $\nu_{B,1}$ comb and emitting into the $\nu_{B,2}$ comb. The phase written to the qubit in this transition is $\Phi_{B,1}-\Phi_{B,2}$, where $\Phi_{B,1}$ and $\Phi_{B,2}$ are the optical phases of the two combs at the ion position. The inverse process from $\ket{1}$ to $\ket{0}$ reverses these phases.  This coherent transition can also be driven directly with microwaves at frequency $\nu_0$.
(b) Simplified experimental diagram. The master microwave oscillator and pulsed laser repetition rate are locked through a feed-forward system.  Acousto-optic modulator (AOM) B is used for copropagating transitions, and AOM A is used in conjunction with AOM B for multi-qubit entangling gates.
}
\label{simplified_circuit}
\end{figure}

The qubit state can be rotated between $\ket{0}$ and $\ket{1}$ with optical or microwave fields, and we demonstrate phase coherence between these operations by using them sequentially on a qubit. Copropagating stimulated Raman transitions~\cite{bible} are driven with optical frequency combs~\cite{comb} generated by a mode-locked 355 nm ($\nu_{PL}\approx 844.48$ THz) pulsed laser with repetition rate $\nu_r$. An acousto-optic modulator (AOM B) is driven with frequencies $\nu_{B,1}$, $\nu_{B,2}$ that are adjusted to bring the beat-note between Raman beams on resonance with the qubit hyperfine splitting~(Fig. \ref{simplified_circuit}a):
\begin{equation}
\begin{split}
&\nu_0 = p\nu_{r} + \nu_{B,1} - \nu_{B,2}
\end{split}
\label{eq:copropagating}
\end{equation}
where $p$ is an integer. Due to atomic selection rules, transitions are only driven when the two beams have the same circular polarization \cite{ybqubit}. Since these beams from AOM B are nominally copropagating, drifts of the optical path length result in negligible phase errors on the qubit.  

In order to stabilize the beat-note frequency to an external master oscillator, we feed-forward fluctuations in the measured repetition rate of the pulsed laser to downstream AOM B (see Fig. \ref{simplified_circuit}b) \cite{phaselock}. This feed-forward technique may be more useful than directly stabilizing the laser cavity length, because of the limited bandwidth of mechanical transducers and the possible inaccessibility of the laser cavity.  

\begin{figure}
\includegraphics[width = 3.3 in]{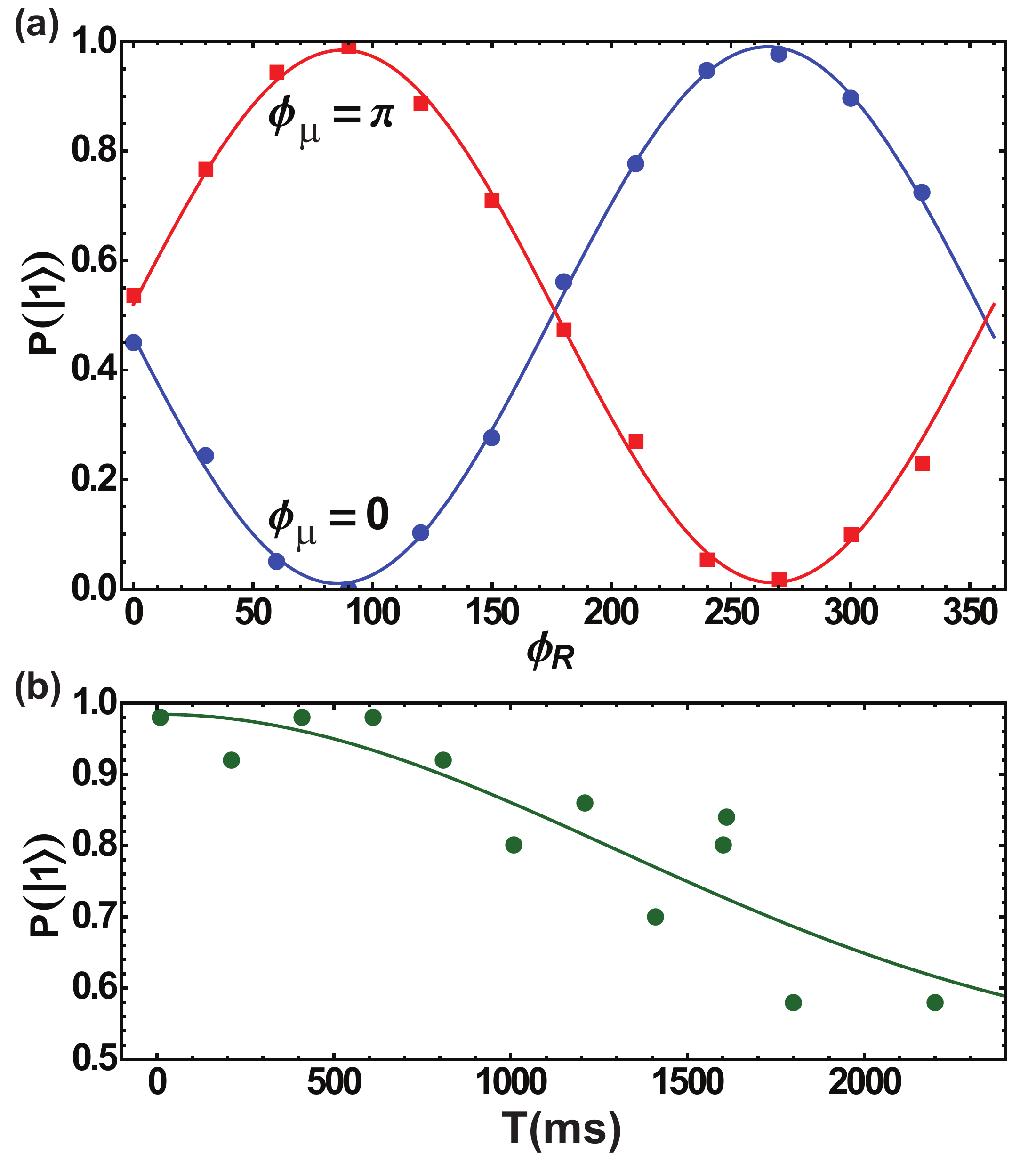}
\caption{(color online). (a) $\pi/2$ microwave rotation followed by a $\pi/2$ Raman rotation. The phase of the Raman rotation, $\phi_{R}$, is scanned for two different microwave phases, $\phi_{\mu} = 0,\pi$. If these operations are phase coherent, the final state of the qubit can be controlled by adjusting the phase of either operation. The probability of being in state $\ket{1}$ is fit to $P(\ket{1})$ = 0.5 + $A$ sin$(\phi_{0}+\phi_{R}-\phi_{\mu})$ where $\phi_{0}$ is a static offset phase that stems from the path length difference between the microwave/optical fields and the ion. (b) A Ramsey experiment with delay $T$ is carried out with an initial Raman rotation and a final microwave rotation. The phases of the rotations are adjusted to give $P(\ket{1})=1$ at $T=0$. A Gaussian fit to the data gives a $1/e$ decay time of 1.8 seconds.}
\label{mw_raman}
\end{figure}

We use the master oscillator as a reference clock for microwave and Raman rotations. In order to maintain phase coherence between qubit operations over long time scales, we use an arbitrary waveform generator (AWG) that provides signals at multiple frequencies with well-defined phase relations and bridges frequency differences between the master oscillator and qubit levels. We verify coherence between microwave and Raman rotations by performing a Ramsey experiment and observe a coherence time of 1.8 seconds as shown in Fig. \ref{mw_raman}. With this scheme, microwaves can be used for global qubit rotations, while focused Raman beams can address individual qubits in a long chain for single qubit rotations.

\subsection{Multi-qubit entangling gates}
Entangling trapped atomic qubits through their Coulomb interaction requires external field gradients that provide state-dependent forces.  The absolute phase and amplitude of microwave or rf fields can easily be controlled for this purpose, but generating sufficiently high field gradients requires specialized trap geometries and high currents \cite{mw_gate}.  Instead, optical fields can be used where non-copropagating Raman beams are required to generate large field gradients \cite{quantum_dynamics, cz, ms}, however, relative path length fluctuations can imprint unknown phases on the qubits.  

Here we utilize a particular geometry of non-copropagating beams to realize gates insensitive to the optical phase of the laser beams. Such gates have been demonstrated on magnetic field sensitive states \cite{sigmaz}; however, their susceptibility to magnetic field noise results in shorter coherence times compared to clock states. Phase insensitive gates on clock states have been realized with CW lasers to provide a state-dependent force by addressing both red and blue sideband transitions; $\ket{0}\ket{n} \rightarrow \ket{1}\ket{n-1}$ and $\ket{0}\ket{n} \rightarrow \ket{1}\ket{n+1}$ respectively where $\ket{n}$ is the vibrational eigenstate of the ions in a harmonic trap potential \cite{bible,cwgate, plee}. This has also been accomplished by simultaneously driving a carrier, $\ket{0}\ket{n} \rightarrow \ket{1}\ket{n}$, and a single sideband transition \cite{carrier_gate_theory, carrier_gate_exp}. However, this approach requires very large carrier Rabi frequencies to prevent additional gate errors \cite{NJP_dressed}. 

The use of CW lasers is technically difficult for systems with qubit splittings more than a few GHz since it requires phase-locking two monochromatic sources or the use of modulators with limited bandwidths. Alternatively, the large bandwidths of ultrafast laser pulses easily spans such splittings \cite{comb}. Here, we experimentally demonstrate a phase insensitive gate on the clock states of two qubits, where two sidebands of a vibrational mode are excited simultaneously by an optical frequency comb generated from a pulsed laser. The beat-note of the frequency combs is locked to the master oscillator to provide phase coherence between quantum gates performed over long time scales and at different locations while maintaining phase coherence of the entangling gates with microwave and copropagating Raman rotations. The techniques demonstrated here can also be used to maintain long coherence times on simultaneous carrier and single sideband gates \cite{carrier_gate_theory}, where the carrier transition is induced either by microwaves or Raman beams. 

\begin{figure*}
\includegraphics[width = 6.6 in]{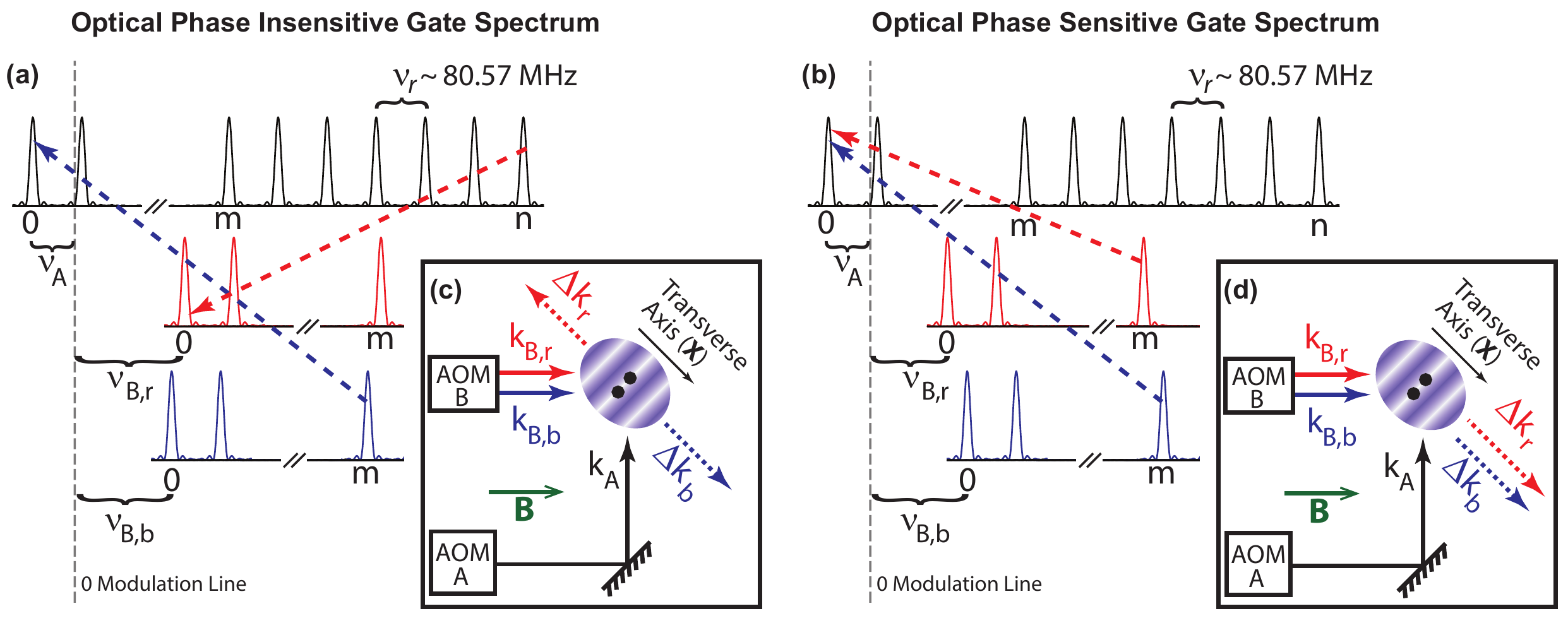}
\caption{(color online). Representation of the optical combs in the frequency domain (a, b) and orientation of the Raman beams with respect to the addressed vibrational mode, $\textbf{X}$, and magnetic field, $\textbf{B}$ (c, d). Beam $\textbf{k}_{A}$ is polarized perpendicular to $\textbf{B}$, while beams $\textbf{k}_{B,r}$ and $\textbf{k}_{B,b}$ have $\sigma_{+}$ polarization. This orientation allows copropagating Raman transitions to be driven by AOM B and the entangling gates to be driven by AOMs A and B. In order to drive the gate, AOMs A and B shift the reference $0^{th}$ comb tooth by $\nu_A$, $\nu_{B,r}$ and $\nu_{B,b}$ from the 0 modulation line (vertical dashed line) and the negative shift for $\nu_A$ is obtained by taking negative first order diffracted beam. The beat-note between the combs, represented by the dashed arrows, have the required frequencies for the gate and the optical field gradient (purple shading) addresses the transverse modes. (a) In the optical phase insensitive geometry, off-resonant blue sideband transition is driven by absorption from the $m^{th}$ comb tooth of the $\textbf{k}_{B,b}$ beam and emission into the $0^{th}$ comb tooth of the $\textbf{k}_{A}$ beam. The absorption and emission directions of the red sideband transition is opposite that of the blue sideband transition such that the gate is driven by absorbing from the $n^{th}$ comb tooth of the $\textbf{k}_{A}$ beam and emitting into the $0^{th}$ comb tooth of the $\textbf{k}_{B,r}$ beam. (b) In the optical phase sensitive geometry, off-resonant red and blue sideband transitions are driven by absorption from the $m^{th}$ comb tooth of the $\textbf{k}_{B,r}$,$\textbf{k}_{B,b}$ beams and emission into the $0^{th}$ comb tooth of the $\textbf{k}_{A}$ beam. (c) In the M{\o}lmer-S{\o}rensen protocol, the gate phase $\phi_{G}=-(\phi_{rsb}+\phi_{bsb})$, where $\phi_{rsb},\phi_{bsb}$ are phases associated with the red and blue sideband transitions. Drifts of the optical path length from the source to the ions, $\delta x$, along the $\textbf{k}_{B,r}$, $\textbf{k}_{B,b}$ beam path change the optical phases of these fields at the ion position resulting in a phase shift of $\phi_{rsb}$ and $\phi_{bsb}$ by $\delta\phi=k_{B,r}\delta x \approx k_{B,b}\delta x$ (see Fig. \ref{simplified_circuit}a and Eq. \ref{eq:spin}). In the optical phase insensitive geometry, since the direction of the red and blue sideband transitions are opposite, the phase changes nearly cancel out so that $\phi_{G}^{'}=(\phi_{rsb}-\delta\phi)+(\phi_{bsb}+\delta\phi)\approx \phi_{G}$, providing optical path length independence to the gate. (d) For the optical phase sensitive case, this change is directly imprinted onto the ions: $\phi_{G}^{'}=(\phi_{rsb}+\delta\phi)+(\phi_{bsb}+\delta\phi) \approx \phi_G+2\delta\phi$. Similar uncorrelated phase sensitivity is also present on path length drifts of the $\textbf{k}_{A}$ beam.}
\label{comb_and_geometry}
\end{figure*}

\subsubsection{Generation of gate frequencies}

Two-qubit entanglement is generated following the M{\o}lmer-S{\o}rensen protocol \cite{ms, Milburn, Solano}, in which optical driving fields are tuned near the red and blue sidebands of a vibrational mode. In order to obtain the desired optical spectra for the phase insensitive gate \cite{comb, phaselock}, each Raman beam passes through AOMs A and B of Fig. \ref{simplified_circuit}b to generate a relative frequency offset ($\nu_A, \nu_{B,r}, \nu_{B,b}$) and allow phase control of the various frequency elements~(Fig.~\ref{comb_and_geometry}a):
\begin{equation}
\begin{split}
&\nu_0-\nu_{\alpha}+\delta = n\nu_{r} - \nu_A - \nu_{B,r}  \\
&\nu_0+\nu_{\alpha}-\delta  = m\nu_{r} + \nu_{B,b} + \nu_A 
\end{split}
\label{eq:combsgate}
\end{equation}
where $n$ and $m$ are integers, $\nu_{\alpha}$ is the frequency of the vibrational mode of interest and $\delta$ is the symmetric detuning from this mode. Note that $\nu_{B,r}$ and $\nu_{B,b}$ are applied to the same AOM, resulting in two nearly copropagating beams. With $\nu_{\alpha} \approx $ 2.5 MHz, $\delta $ = 10 kHz, $\nu_{r} \approx$ 80.57 MHz and $\nu_A$ = 77.5 MHz, these equations can be satisfied by $n=160$, $\nu_{B,r} \approx$ 173.4 MHz and $m=154$,  $\nu_{B,b} \approx$ 160.0 MHz.

\subsubsection{The gate phase}

After application of the optical fields for the gate time, the collective motion of the ions factors and the qubit states evolve as \cite{ms, plee}:
\begin{equation}
\begin{split}
\ket{00} &\rightarrow  \ket{00} -ie^{-i\phi_{G}}\ket{11} \\
\ket{11} &\rightarrow \ket{11} -ie^{i\phi_{G}}\ket{00} \\
\end{split}
\quad \quad
\begin{split}
\ket{01} &\rightarrow  \ket{01}-i\ket{10} \\
\ket{10} &\rightarrow  \ket{10}-i\ket{01}
\end{split}
\label{eqn:molmer}
\end{equation}
The gate phase is $\phi_{G}=\phi_{S,i}+\phi_{S,j}$ with individual ``spin" phases: 
\begin{equation}
\begin{split}
\phi_{S,i}&=-(\phi_{rsb,i}+\phi_{bsb,i}) \\
&=-\frac{1}{2}(\Delta{\textbf{k}}_{r} \boldsymbol{\cdot} \textbf{X}_{i}-\Delta{\phi_{r}}+\Delta{\textbf{k}}_{b}\boldsymbol{\cdot}\textbf{X}_{i}-\Delta{\phi_{b}}).
\end{split}
\label{eq:spin}
\end{equation}
Here $\phi_{rsb,i}$, $\phi_{bsb,i}$ are the phases associated with the red and blue sideband transitions and $\textbf{X}_{i}$ is the position of the $i^{th}$ ion \cite{plee}. The two optical field pairs address the red ($\textbf{k}_{A}$, $\textbf{k}_{B,r}$) and blue ($\textbf{k}_{A}$, $\textbf{k}_{B,b}$) vibrational sidebands. To drive the red sideband using a mode-locked pulsed laser, a photon is absorbed from the $\textbf{k}_{A}$ comb tooth and emitted into the $\textbf{k}_{B,r}$ comb tooth. The opposite process takes place for the blue sideband, resulting in $\Delta{\textbf{k}}_{r} = \textbf{k}_{A} - \textbf{k}_{B,r}$ and $\Delta{\textbf{k}}_{b}=\textbf{k}_{B,b} - \textbf{k}_{A}$. Since the $\Delta{\textbf{k}}$ vectors point in opposite directions, $\Delta{\textbf{k}}_{r} \approx -\Delta{\textbf{k}}_{b}$, small fluctuations of the optical path length cancel to a high degree, leaving the gate phase unchanged (Fig. \ref{comb_and_geometry}c,d). The gate phase retains sensitivity to the rf signals applied to the AOMs and may be modified by modulating the applied phases $\phi_A,\phi_{B,r}$ and $\phi_{B,b}$ to set $\Delta{\phi_{r}}=\phi_{A}-\phi_{B,r}$ and $\Delta{\phi_{b}}=\phi_{B,b}-\phi_{A}$ to any desired value.

\subsubsection{The motional phase}

During an entangling gate, the motion correlated with particular eigenstates of the two qubits are separated in phase space with application of a state-dependent force.  Without loss of generality, we consider a single collective mode of motion, and the relative displacements are described by the motional phase \cite{plee}
\begin{equation}
\phi_{M,i}=\frac{1}{2}\left(\Delta{\textbf{k}}_{r}\boldsymbol{\cdot}\textbf{X}_{i}
                   -\Delta{\phi_{r}}-\Delta{\textbf{k}}_{b}\boldsymbol{\cdot}\textbf{X}_{i}+\Delta{\phi_{b}}\right). 
\end{equation}
In the optical ``phase insensitive" geometry \cite{plee}, the optical path length dependence of $\phi_{S,i}$ is transferred to $\phi_{M,i}$; however, the phase dependence of $\phi_{M,i}$ on the optical path is identical for the two ions and thus global fluctuations do not affect the entangling gate \cite{motional_phase}. 

\begin{figure}
\includegraphics[width = 3.3 in]{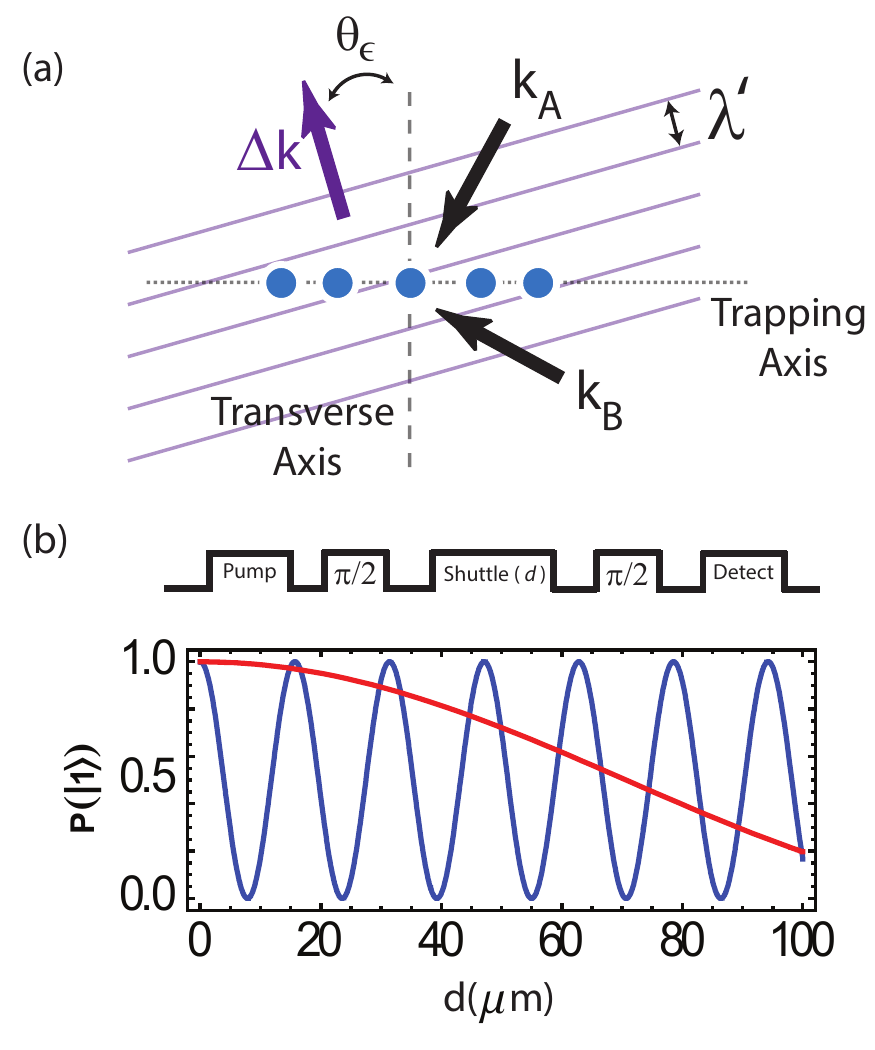}
\caption{(color online). (a) If the wave fronts of the optical field gradient (purple lines) are misaligned with respect to the trapping axis by an angle $\theta_{\epsilon}$, the ions experience different state-dependent force phases resulting in gate errors. The wave fronts are separated by $\lambda^{'}=\frac{2\pi}{\Delta{k}}\approx\frac{2\pi}{\sqrt{2}{k}}\approx$ 250 nm. As an example, in order to realize a phase variation of $\textless$10$^\circ$ along a 30 $\mu$m ion chain, $\theta_{\epsilon}$ must be $\textless$0.02$^\circ$. (b) Experimental sequence for wave front alignment and expected signal. A single ion in the state $\ket{0}$ is rotated by a resonant non-copropagating Raman $\pi/2$ pulse and shuttled by $d$ along the trapping axis. In the new position, the ion is rotated again by another non-copropagating Raman $\pi/2$ pulse before fluorescent detection of the final state. The blue (red) curve shows the expected ion brightness corresponding to a 1$^\circ$ (0.05$^\circ$) misalignment. The oscillation on the final qubit state is a result of the phase difference between the resonant $\pi/2$ rotations and is given by $P(\ket{1})=$cos$^{2}(\pi d $sin$(\theta_{\epsilon})/\lambda^{'})$.}
\label{misalignmnet}
\end{figure}

The static motional phase difference between two ions $\phi_{M_i}-\phi_{M_j}$ determines the gate time \cite{plee} to produce the evolution of Eq. \ref{eqn:molmer}. If axial vibrational modes are used, the distance between the ions must be carefully controlled and the gate fidelity becomes susceptible to changes in ion spacing \cite{cwgate,carrier_gate_exp}. Moreover, entangling longer ion chains becomes problematic as the distance between ions may vary along the chain. These issues are circumvented by using the transverse modes for gate operations \cite{transverse_modes}. Since the phase fronts created by the optical fields are ideally uniform across the trapping axis when the transverse modes are addressed, the motional phase is the same for all ions (Fig. \ref{comb_and_geometry}c,d). However, misalignment between the $\Delta{\textbf{k}}$ vectors and the transverse axis by an angle $\theta_{\epsilon}$ would introduce a motional phase difference $\Delta{\phi_{\epsilon}}=\Delta{k}l$sin$(\theta_{\epsilon})$ between the ions where $l$ is the ion separation (Fig. \ref{misalignmnet}a).

Optical fields can be aligned to better than $\theta_{\epsilon}$ $\textless$ 0.05$^\circ$ by measuring the variation of the resonant photon scattering rate across the ions due to the AC Stark shift induced by the optical field gradient \cite{penning_alignment}. Since this technique relies on obtaining sufficiently large AC Stark shifts, it requires tuning the Raman beam frequencies close to the Doppler cooling transition which may be impractical with pulsed lasers due to their large bandwidths and limited tuning capabilities. Furthermore, achieving good alignment relies on using large ion crystals; while an ion crystal diameter of hundreds of $\mu$m can be maintained in Penning traps \cite{penning_alignment}, it can be challenging to hold similar length ion crystals in rf Paul traps. An alternative technique incorporates shuttling and utilizes the phase differences of non-copropagating Raman rotations at different points along the trapping axis. The phase differences could be directly measured using a single ion for the alignment of the Raman beams with respect to the transverse axis (see Fig. \ref{misalignmnet}b). Although not implemented in this work, high accuracy alignment can be achieved in principle with this technique.

\begin{figure}
\includegraphics[width = 3.3 in]{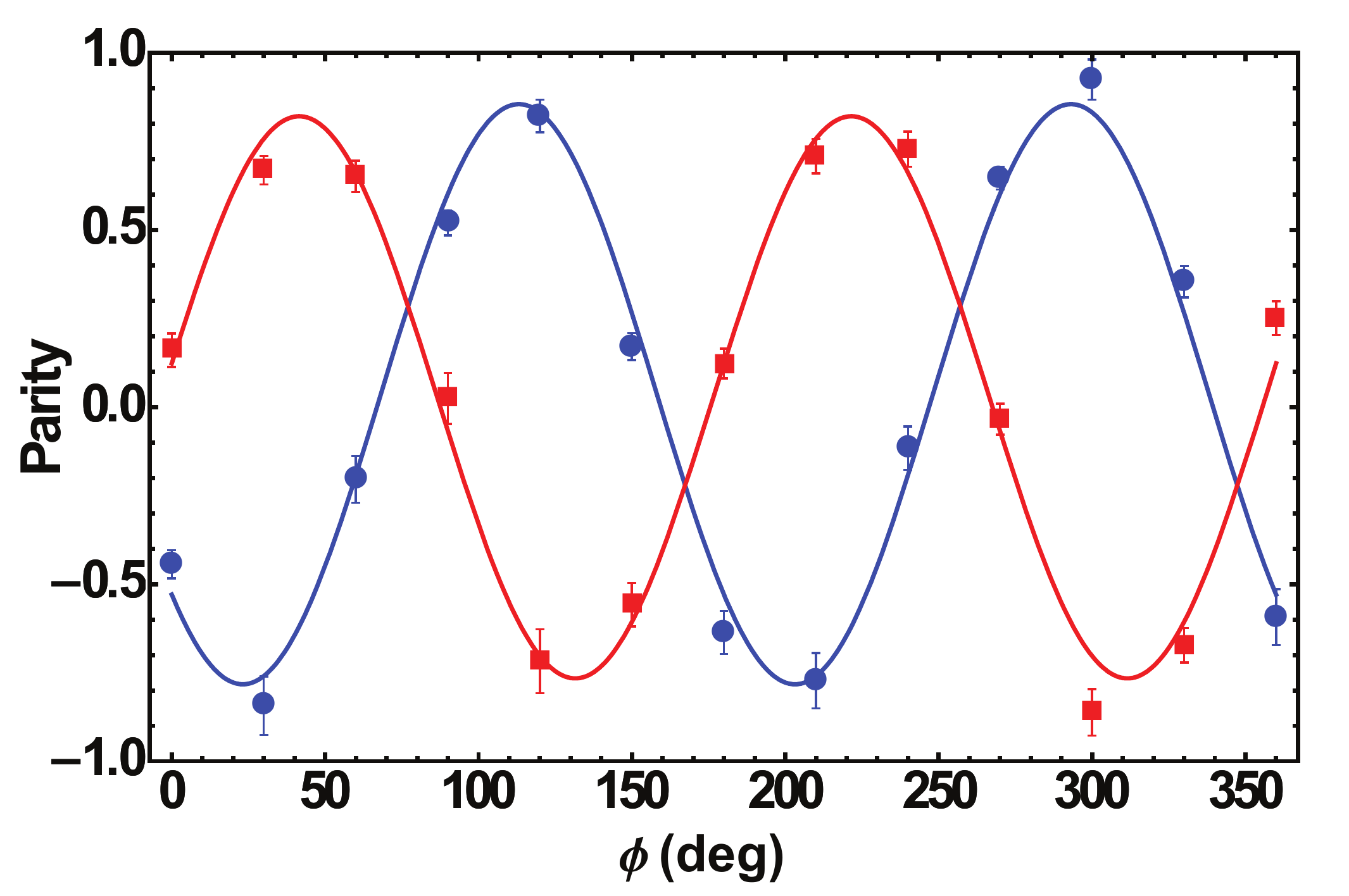}
\caption{(color online). Parity, $P(\ket{00}) + P(\ket{11}) - P(\ket{01}) - P(\ket{10}) = A$ cos$(\phi_{G}+2\phi+\phi^{'})$ , of the two qubit entangled state. Ions are first optically pumped to the $\ket{00}$ state and following the phase insensitive gate, a $\pi/2$ analysis rotation with phase $\phi$ is applied. Blue circles are the result of analysis with a copropagating Raman rotation and red squares are analyzed with a microwave rotation. The phase shift between the parity curves is due to different $\phi^{'}$ static offsets between the gate and the $\pi/2$ analysis rotations.}
\label{gate_analyzed}
\end{figure}  

\subsubsection{Phase coherence of the gate}

Long term phase coherence can be maintained with an extension of the beat-note stabilization technique by feeding forward changes in $\nu_{r}$ to $\nu_{B,r},\nu_{B,b}$ (see Appendix for details). Even in the absence of drifts in $\nu_{r}$, this technique can be used to synchronize pulsed laser operations with a master oscillator to maintain phase coherence with microwaves or operations by other pulsed lasers in the system. A free-running frequency source can be used to generate the AOM frequency $\nu_A$ as $\phi_{A}$ cancels in the gate phase, $\phi_{G}=\Delta{\phi_{r}}+\Delta{\phi_{b}}=(\phi_{A}-\phi_{B,r})+(\phi_{B,b}-\phi_{A})$. In order to maintain phase coherence between entangling gates, copropagating Raman transitions and microwave rotations that have differing drive frequencies, an AWG may be used for these operations rather than free-running frequency sources, where phase relations between different frequency components must be tracked resulting in increased system overhead.

\subsubsection{Characterization of the system}

We characterize the optical phase sensitivity of entangling gates by measuring the fidelity of various entangled states through extraction of the density matrix elements of the prepared state \cite{fidelity}; we measure the populations along with the parity contrast in order to extract a fidelity of $\mathcal{F}$ $\approx$ 0.86. The parity contrast is obtained by scanning the phase of the analysis microwave and Raman $\pi/2$ pulses after the entangling gate (Fig. \ref{gate_analyzed}). For the gate, Walsh modulation is implemented to suppress detuning and timing errors \cite{walsh}. The imperfect fidelity is not a limitation of the phase insensitive gate; we observe similar fidelities using a phase sensitive geometry (Fig.~\ref{comb_and_geometry}b,d) for the gate. Thermal populations of the motional states contributes an error of $\sim8$\% and histogram fitting of two ion combined brightness for parity measurements contributes an additional $\sim5$\% \cite{comb}.

\begin{figure}
\includegraphics[width = 3.3 in]{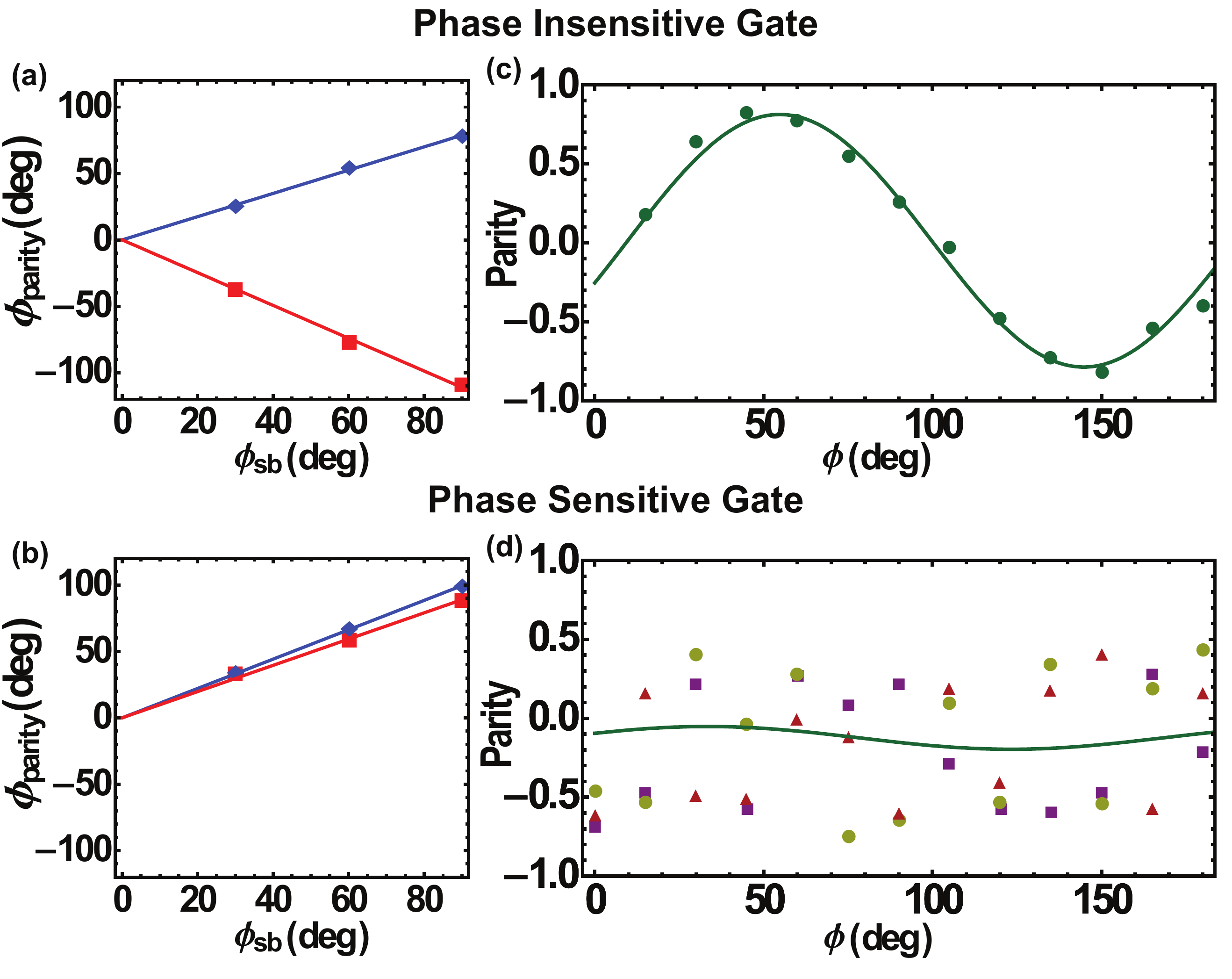}
\caption{(color online). (a)-(b) Changes in the phase of the red ($\phi_{sb} \equiv \phi_{B,r}$) and blue ($\phi_{sb} \equiv \phi_{B,b}$) sideband addressing frequencies cause $\phi_{G}$ to shift in opposite directions for the phase insensitive gate while $\phi_{G}$ shifts in the same direction for the phase sensitive gate. This behavior is verified by the phase shift of the parity oscillation. (c)-(d) To simulate a change in the relative optical path length, a random phase is added to frequencies provided by the AWG during the gate at each point. The parity curve is not affected for the phase insensitive gate, while the phase sensitive gate parity curve becomes randomized from point to point as verified by three data sets.}
\label{sideband_movement}
\end{figure}

We further characterize and compare the phase insensitive and sensitive gates by directly measuring how the phases of the driving fields are imprinted on the entangled states. In the case of a phase insensitive gate, the phase of the red and blue sideband frequencies modify the gate phase with opposite signs, $\phi_{G} \approx \phi_{B,b}-\phi_{B,r}$. The phase of the parity oscillation shift in opposite directions for red and blue sideband phase shifts. In the phase sensitive case, $\phi_{G} \approx \phi_{B,r}+\phi_{B,b}-2\phi_{A}$, which results in the parity phase moving in the same direction for both sideband phase shifts (Fig \ref{sideband_movement}a,b). To simulate a relative optical path length change at the ion position, a random phase is added to both sidebands driven by the AWG. The phase insensitive gate parity is not affected by this randomization process, while loss of contrast is observed for the phase sensitive gate as expected (Fig~\ref{sideband_movement}c,d). 

Lastly, we test the stability of our system over long time scales by monitoring the phase of parity oscillations following analysis of the phase insensitive gate by a microwave pulse. We observe phase fluctuations of \textless 8$^\circ$ of the parity curve over a period of 24 hours. Therefore, once relative phase relations have been characterized between different quantum operations sharing the same master oscillator, regular monitoring of these phases is not necessary. This long term stability will be necessary for long computations.

\section{OUTLOOK: IMPLICATIONS TOWARDS SCALABILITY}
The techniques presented here can be useful in a large scale modular quantum processor architecture \cite{musiqc,modular}. In this proposal, modules hold ion chains of manageable sizes and entanglement within a module is generated with mutual Coulomb interactions while photonic interfaces \cite{remote_entanglement_theory, moehring} establish connections between separate modules. As shown here, the use of a common master oscillator for all quantum operations and insensitivity to optical path length fluctuations can be implemented to realize phase coherent operations across this architecture.

In the shuttling model proposed for a large-scale quantum processor, ions are transported between various trapping regions in order to perform specific operations \cite{large_scale}. These phase stabilization techniques might be beneficial in this model as it is important to maintain phase coherence between the operations performed at different regions of the processor and at different times. Moreover, coupling to transverse modes for multi-qubit gate operations instead of axial modes would eliminate errors that might stem from small changes in ion separation after shutting between regions. Finally, the complexity of the device electrode structure might be reduced as it is not necessary to keep a uniform ion spacing with the use of transverse modes \cite{transverse_modes}. 

\begin{figure*}
\includegraphics[width = 6.6 in]{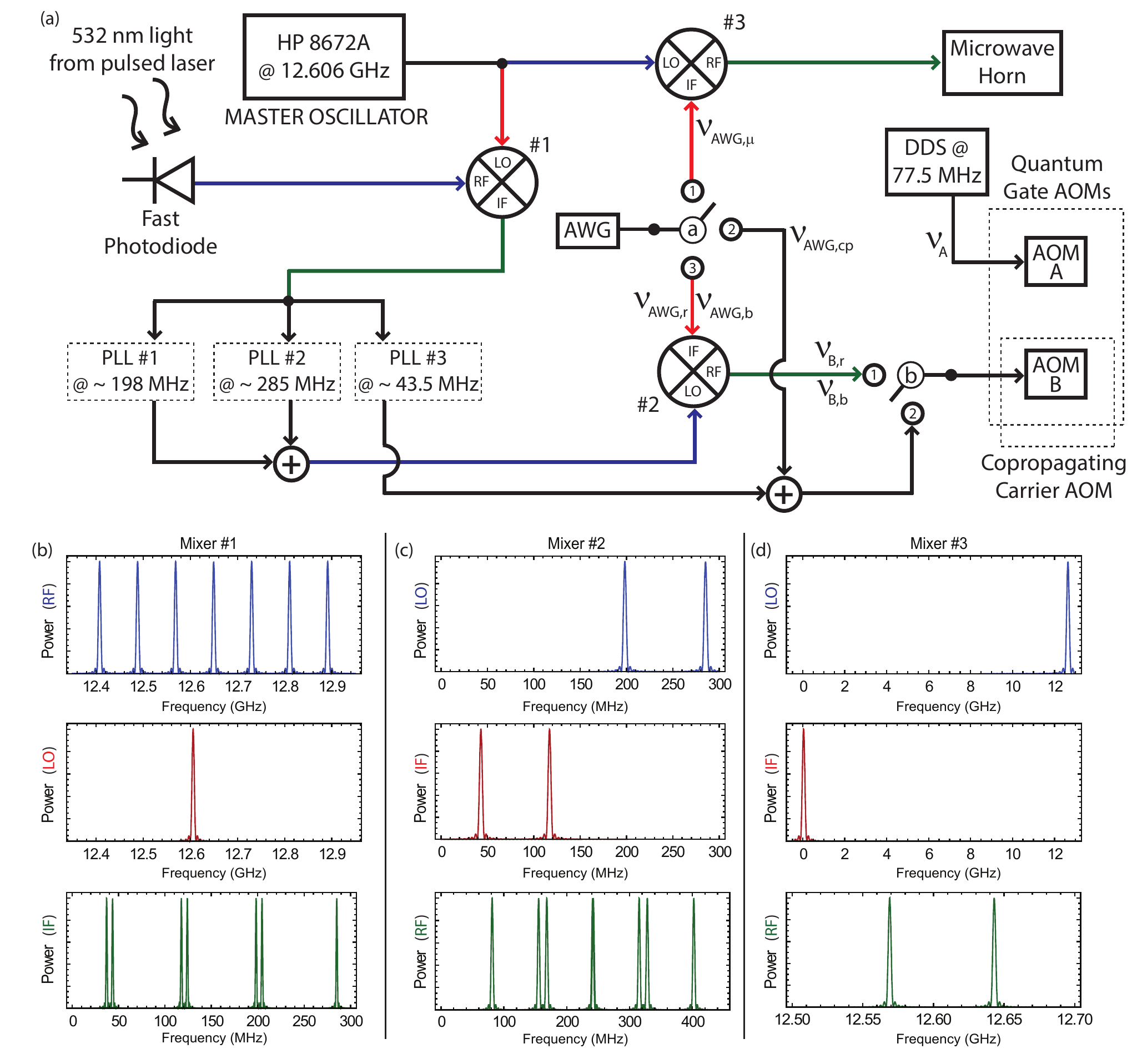}
\caption{(color online). (a) Phase coherence circuit. An AWG is used to provide the necessary frequencies for quantum operations while at the same time maintaining phase relations between different frequency components. Filters are used throughout the circuit to remove undesired frequency components of the mixer output. The second harmonic light of a mode locked Nd:YAG laser at 532 nm is directed to a fast photodiode which generates a frequency comb with tooth separation $\nu_{r}$. The third harmonic at 355 nm is used to drive atomic transitions. (b) The photodiode signal is mixed ($\#1$) with the master oscillator (HP 8672A) and sent to three different PLLs which use this signal to output $\sim 198$ MHz and $\sim 285$ MHz, matching the difference between the oscillator and the $m=154$, $n=160$ comb teeth. (c) The PLL 1 and 2 signals are first combined and then mixed ($\#2$) with the AWG to address the detuned sideband frequencies of the trapped ions. During the gate, switch a $\rightarrow$ 3 and switch b $\rightarrow$ 1. (d) Phase coherent microwave rotations with gates are realized by mixing ($\#3$) the AWG signal with the master oscillator to drive carrier transitions. For the microwave rotations, switch a $\rightarrow$ 1. The third PLL provides phase coherent copropagating carrier transitions using the $p=157$ comb tooth and AOM B, with switch a $\rightarrow$ 2 and switch b $\rightarrow$ 2.}
\label{lock_circuit_real}
\end{figure*}

\begin{figure*}
\includegraphics[width = 6.6 in]{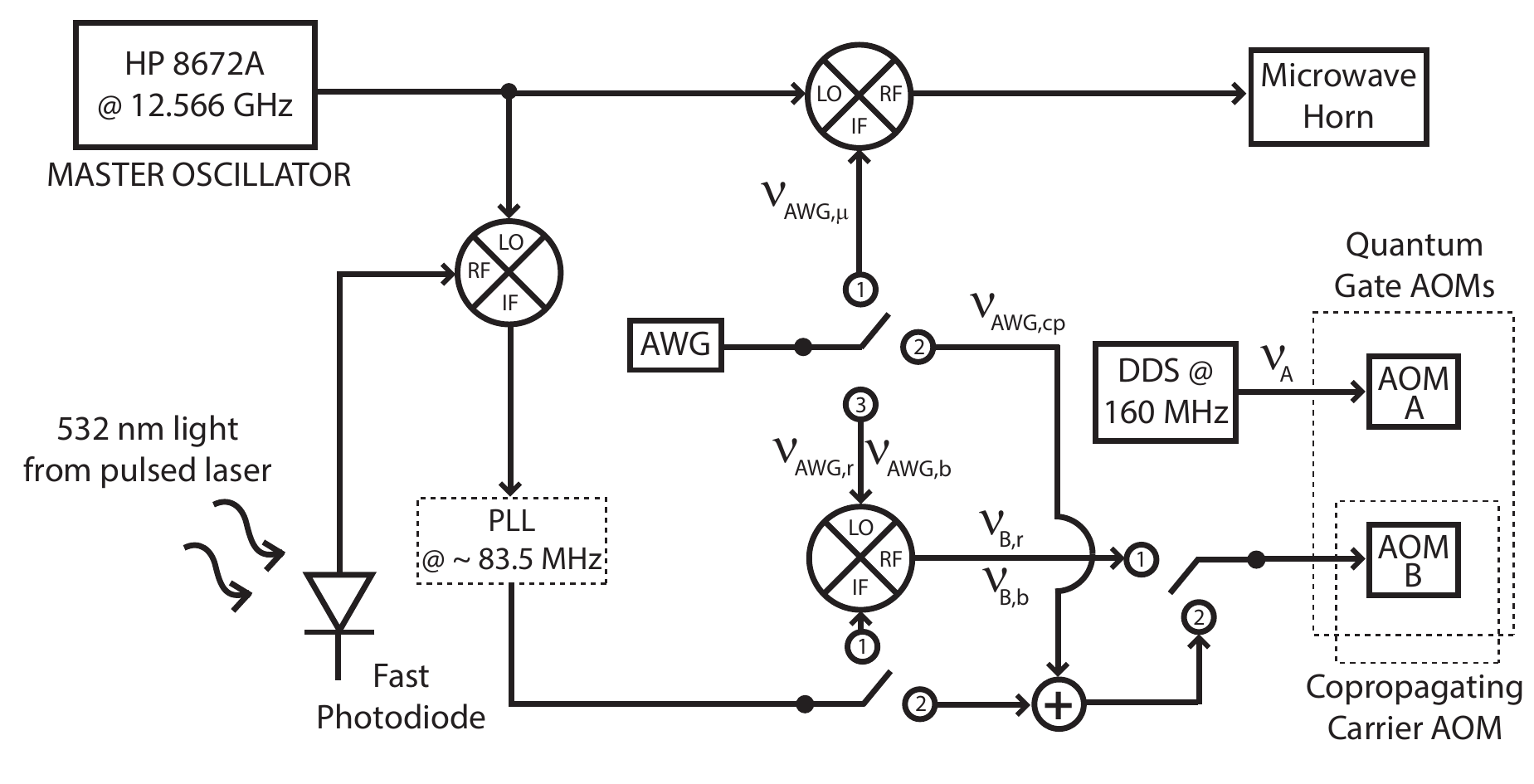}
\caption{Phase coherent circuit with single PLL for phase coherent qubit operations.}
\label{lock_circuit_ideal}
\end{figure*}

\section{SUMMARY}
In summary, we demonstrate long term coherence between various qubit operations utilizing optical and microwave fields referenced to a single master oscillator. The setup presented here effectively eliminates any optical path length related phase drifts from these operations, obviating the need for optical interferometric stability in a quantum system. Moreover, the use of a master oscillator as a reference provides coherence between qubit operations done at different times and at different locations which is central to realizing a large-scale, distributed and modular quantum computer. By using a stable master oscillator, the long coherence times of trapped atomic ions can be harnessed effectively to execute many subsequent operations on the system and preserve quantum information for long times while operations are performed on other qubits. 
 
\begin{acknowledgments}
This work was supported by the Intelligence Advanced Research Projects Activity, the Army Research Office MURI Program on Hybrid Quantum Optical Circuits, Defense Advanced Research Projects Agency SPARQC and the NSF Physics Frontier Center at JQI.
\end{acknowledgments}

\appendix*
\section{Phase Stabilization Circuit}

Cavity length changes cause drifts in the repetition rate of the pulsed laser,  $\nu_{r} + \delta_{r}(t)$, which result in fluctuations of the separation between comb teeth (Fig. \ref{comb_and_geometry}a,b) and thus phase and frequency drifts that can cause gate errors. Since two different comb tooth solutions are used to drive the gate (Eq. \ref{eq:combsgate}), separate phase locked loops (PLLs) are necessary to lock the $m\nu_r$ and $n\nu_r$ frequency splittings between the comb teeth (see Ref. \cite{phaselock} for details on the PLL). Moreover, phase coherence between quantum operations is needed for full qubit control and can be achieved with the circuit given in Fig. \ref{lock_circuit_real}. By adding a third PLL, coherent copropagating Raman carrier transitions can also be incorporated.  

In order to monitor and feed-forward the repetition rate drift $\delta_{r}(t)$, the signal from the fast photodiode is mixed with the master oscillator, $\nu_{MO}=12.606$ GHz, to produce beat-notes. The PLLs output a signal that is phase locked with the relevant input beat-note frequencies:
\begin{equation}
\begin{split}
&\nu_{PLL1}=n[\nu_{r} + \delta_{r}(t)]-\nu_{MO} \\
&\nu_{PLL2}=\nu_{MO}-m[\nu_{r} + \delta_{r}(t)]
\end{split}
\label{eq:PLLs}
\end{equation}
where $m=154$ and $n=160$ in this experiment. These output signals are mixed with the AWG signal to provide driving frequencies for AOM B:
\begin{equation}
\begin{split}
&\nu_{B,r}=\nu_{PLL1}-\nu_{AWG,r} \\
&\nu_{B,b}=\nu_{PLL2}-\nu_{AWG,b}
\end{split}
\label{eq:AOM2}
\end{equation}
Both frequencies should be within the bandwidth of AOM B for optimal diffraction efficiency. Inserting Eq. \ref{eq:PLLs} and \ref{eq:AOM2} in Eq. \ref{eq:combsgate} with $\nu_{A} =~77.5$ MHz, the AWG frequencies for driving the entangling gate are:
\begin{equation}
\begin{split}
\nu_{AWG,r}&=\nu_{PLL1}-n[\nu_{r} + \delta_{r}(t)]+ \nu_{A} +  \nu_{0} - \nu_{\alpha} + \delta  \\
&=-\nu_{MO} + \nu_{A} +  \nu_{0} - \nu_{\alpha} + \delta\\
\nu_{AWG,b}&=\nu_{PLL2}+m[\nu_{r} + \delta_{r}(t)]+ \nu_{A} -  \nu_{0}  - \nu_{\alpha} + \delta\\
&=\nu_{MO} + \nu_{A} -  \nu_{0} + \nu_{\alpha} + \delta\\
\end{split}
\label{eq:AWGs}
\end{equation}
with $\nu_{AWG,r} \approx 116.8$ MHz, $\nu_{AWG,b} \approx 43.2$ MHz. As can be seen from Eq. \ref{eq:AWGs}, feed-forward to the PLLs not only eliminates sensitivity to $\delta_{r}(t)$ but also utilizes the master oscillator $\nu_{MO}$ as a reference for qubit transitions. To generate microwave rotations that are phase coherent with the Raman transitions, the master oscillator is mixed with the AWG, $\nu_{AWG,\mu}=\nu_{0}-\nu_{MO}$, and sent to a microwave horn. The achievable coherence time between quantum operations with this technique can be increased by using oscillators with lower phase noise. 

It is also possible to realize the set of operations presented in this paper by using only one comb tooth solution, $n=m=157$, with $\nu_A = 160$ MHz, $\nu_{B,r} \approx$ 169.2 MHz and $\nu_{B,b} \approx$ 155.7 MHz. Through the appropriate use of mixers, a single PLL can provide the correct feed-forward to lock these two Raman transitions to the master oscillator (Fig. \ref{lock_circuit_ideal}). This approach has the advantage of using fewer electronic elements.

In Fig. \ref{lock_circuit_real} and \ref{lock_circuit_ideal}, AOM B is used for both entangling gates and copropagating Raman rotations for optimal use of resources. Since the AOMs only work efficiently in a certain rf range, conversion of the rf signals might be necessary to obtain high efficiency beam diffraction for the copropagating Raman rotations. This can be achieved by mixing the rf signals with a DDS to convert signals to the correct frequency range (not shown in Fig. \ref{lock_circuit_real} and \ref{lock_circuit_ideal} for simplicity). As this mixing will result in a common-mode phase and frequency change in both AWG and PLL signals, the DDS signal has no effect on the phase of the rotations hence a free-running source can be used.

\bibliography{paperrefs}

\end{document}